\input{aipcheck}

\documentclass[
    ,final            
    ,draft            
    ,numberedheadings 
  ]
  {aipproc}

\usepackage{amsmath}
\usepackage{amssymb}

\layoutstyle{6x9}

\begin{document}

\title{\textbf{Modified Gravity Or Dark Matter?}}

\author{J. W. Moffat\footnote{Talk given at the International Conference on Two Cosmological Models, Universidad Iberoamericana, Ciudad de Mexico, November 17-19, 2010.}}{address={Perimeter Institute for Theoretical Physics, Waterloo, Ontario N2L 2Y5, Canada\\Department of Physics, University of Waterloo, Waterloo, Ontario N2L 3G1, Canada}}

\begin{abstract}
Modified Gravity (MOG) has been used successfully to explain the rotation curves of galaxies, the motion of galaxy clusters, the Bullet Cluster, and cosmological observations without the use of dark matter or Einstein's cosmological constant. We review the main theoretical ideas and applications of the theory to astrophysical and cosmological data.
\end{abstract}

\maketitle

\section{Introduction}

The ingredients of the standard model of astrophysics and cosmology are:
\begin{enumerate}

 \item General Relativity,

 \item Large-scale homogeneity and isotropy,

 \item 5\% ordinary matter (baryons and electrons),

 \item 25\% dark matter,

 \item 70\% dark energy,

 \item Uniform CMB radiation, $T\sim 2.73$ degrees,

 \item Scale-free adiabatic fluctuations $\Delta T/T\sim 10^{-5}$.

\end{enumerate}

Although the model fits available astrophysical and cosmological data \cite{Komatsu}, it opens up the mystery that about 95\% of all matter and energy are invisible. The dark matter (DM) does not interact with ordinary baryonic matter and light. No current experiments have succeeded in detecting DM . The SN1a supernovae data \cite{Perlmutter,Riess} have created the need for the expansion of the universe to accelerate, promoting the mechanism of dark energy.

In the event that DM is not detected, then to fit all available astrophysical and cosmological data, we are required to modify Newtonian and Einstein gravity without assuming the undetected DM. A modified gravity (MOG) theory, also known as Scalar-Tensor-Vector Gravity or STVG \cite{Moffat,Toth,Toth2}, is based on an action that incorporates, in addition to the Einstein-Hilbert term and the matter action, a massive vector field, and three scalar fields corresponding to running values of the gravitational constant, the vector field coupling constant, and the vector field mass.

We begin in Section~\ref{sec:theory} by introducing the theory through the action principle, and establish key assumptions that allow us to analyze physically relevant scenarios. In Section~\ref{sec:fldeqs}, we derive the field equations using the variational principle. In Section~\ref{sec:static} we solve the field equations in the static, spherically symmetric case. In Section~\ref{sec:testpart}, we postulate the action for a test particle, and obtain approximate solutions to the field equations for a spherically symmetric gravitational field. In Section~\ref{sec:cosmo} we demonstrate how the Friedmann equations of cosmology can be obtained from the theory. In Section~\ref{sec:obs}, we utilize the theory to obtain estimates for galaxy rotation curves, galaxy cluster dynamics and show how the solutions we obtained for the field equations remain valid from cosmological to solar system scales. Lastly, we end in Section~\ref{sec:concl} with conclusions.

\section{Modified Gravity Theory}
\label{sec:theory}

The action of our theory is constructed as follows \cite{Moffat}. We start with the Einstein-Hilbert Lagrangian density that describes the geometry of spacetime:
\begin{equation}
{\cal L}_G=-\frac{1}{16\pi G}\left(R+2\Lambda\right)\sqrt{-g},
\end{equation}
where $G$ is the gravitational constant, $g$ is the determinant of the metric tensor $g_{\mu\nu}$ (we are using the metric signature $(+,-,-,-)$), and $\Lambda$ is the cosmological constant. We set the speed of light, $c=1$. The Ricci-tensor is defined as
\begin{equation}
R_{\mu\nu}=\partial_\alpha\Gamma^\alpha_{\mu\nu}-\partial_\nu\Gamma^\alpha_{\mu\alpha}+\Gamma^\alpha_{\mu\nu}\Gamma^\beta_{\alpha\beta}
-\Gamma^\alpha_{\mu\beta}\Gamma^\beta_{\alpha\nu},
\end{equation}
where $\Gamma^\alpha_{\mu\nu}$ is the Christoffel-symbol, while $R=g^{\mu\nu}R_{\mu\nu}$.

We introduce a ``fifth force'' vector field $\phi_\mu$ via the Maxwell-Proca Lagrangian density:
\begin{equation}
{\cal L}_\phi=-\frac{1}{4\pi}\omega\left[\frac{1}{4}B^{\mu\nu}B_{\mu\nu}-\frac{1}{2}\mu^2\phi_\mu\phi^\mu+V_\phi(\phi)\right]\sqrt{-g},
\end{equation}
where $B_{\mu\nu}=\partial_\mu\phi_\nu-\partial_\nu\phi_\mu$, $\mu$ is the mass of the vector field, $\omega$ characterizes the strength of the coupling between the ``fifth force'' and matter, and $V_\phi$ is a self-interaction potential.

Next, we promote the three constants of the theory, $G$, $\mu,\omega$, to scalar fields by introducing associated kinetic and potential terms in the Lagrangian density:
\begin{eqnarray}
{\cal L}_S&=&-\frac{1}{G}\left[\frac{1}{2}g^{\mu\nu}\left(\frac{\nabla_\mu G\nabla_\nu G}{G^2}+\frac{\nabla_\mu\mu\nabla_\nu\mu}{\mu^2}-\nabla_\mu\omega\nabla_\nu\omega\right)\right.\nonumber\\
&&\left.+\frac{V_G(G)}{G^2}+\frac{V_\mu(\mu)}{\mu^2}+V_\omega(\omega)\right]\sqrt{-g},
\end{eqnarray}
where $\nabla_\mu$ denotes covariant differentiation with respect to the metric $g_{\mu\nu}$, while $V_G$, $V_\mu$, and $V_\omega$ are the self-interaction potentials associated with the scalar fields.

Our action integral takes the form
\begin{equation}
S=\int{({\cal L}_G+{\cal L}_\phi+{\cal L}_S+{\cal L}_M)}~d^4x,
\label{eq:FldL}
\end{equation}
where ${\cal L}_M$ is the ordinary matter Lagrangian density, such that the energy-momentum tensor of matter takes the form:
\begin{equation}
T_{\mu\nu}=-\frac{2}{\sqrt{-g}}\frac{\delta S_M}{\delta g^{\mu\nu}},
\end{equation}
where $S_M=\int{\cal L}_M~d^4x$. A ``fifth force'' matter current can be defined as:
\begin{equation}
J^\nu=-\frac{1}{\sqrt{-g}}\frac{\delta S_M}{\delta\phi_\nu}.
\end{equation}

We assume that the variation of the matter action with respect to the scalar fields vanishes:
\begin{equation}
\frac{\delta S_M}{\delta X}=0,
\end{equation}
where $X=G,\mu,\omega$.

\section{Field equations}
\label{sec:fldeqs}

The field equations of the theory can be obtained in the form of the first and second-order Euler-Lagrange equations~\cite{Toth}:
\begin{equation}
\frac{1}{4\pi}\left[\omega\nabla_\mu B^{\mu\nu}+\nabla_\mu\omega B^{\mu\nu}+\omega\mu^2\phi^\nu-\omega\frac{\partial V_\phi(\phi)}{\partial\phi_\nu}\right]=J^\nu,
\end{equation}
\begin{equation}
\nabla^\nu\nabla_\nu\mu-\frac{\nabla^\nu\mu\nabla_\nu\mu}{\mu}-\frac{\nabla^\nu G\nabla_\nu\mu}{G}+\frac{1}{4\pi}G\omega\mu^3\phi_\mu\phi^\mu+\frac{2}{\mu}V_\mu(\mu)-V'_\mu(\mu)=0,
\end{equation}
\begin{eqnarray}
\nabla^\nu\nabla_\nu\omega-\frac{\nabla^\nu G\nabla_\nu\omega}{G}-\frac{1}{8\pi}G\mu^2\phi_\mu\phi^\mu+\frac{G}{16\pi}B^{\mu\nu}B_{\mu\nu}+\frac{1}{4\pi}GV_\phi(\phi)
\nonumber\\
+V'_\omega(\omega)=0,
\end{eqnarray}
\begin{eqnarray}
\nabla^\nu\nabla_\nu G-\frac{3}{2}\frac{\nabla^\nu G\nabla_\nu G}{G}+\frac{G}{2}\left(\frac{\nabla^\nu\mu\nabla_\nu\mu}{\mu^2}-\nabla^\nu\omega\nabla_\nu\omega\right)
+\frac{3}{G}V_G(G)\nonumber\\
-V'_G(G)+G\left[\frac{V_\mu(\mu)}{\mu^2}+V_\omega(\omega)\right]+\frac{G}{16\pi}(R+2\Lambda)=0,
\end{eqnarray}
\begin{eqnarray}
\left(\frac{2\nabla_\alpha G\nabla_\beta G}{G^2}-\frac{\nabla_\alpha\nabla_\beta G}{G}\right)(g^{\alpha\beta}g_{\mu\nu}-\delta^\alpha_\mu\delta^\beta_\nu)
\nonumber\\
-8\pi\left[\left(\frac{1}{4\pi}G\omega\mu^2\phi_\alpha\phi_\beta-\frac{\partial_\alpha G\partial_\beta G}{G^2}-\frac{\partial_\alpha\mu\partial_\beta\mu}{\mu^2}+\partial_\alpha\omega\partial_\beta\omega\right)\right.
\nonumber\\
\times\left(\delta^\alpha_\mu\delta^\beta_\nu-\frac{1}{2}g^{\alpha\beta}g_{\mu\nu}\right)
\nonumber\\
\left.+\frac{1}{4\pi}G\omega\left(B^\alpha{}_\mu B_{\nu\alpha}+\frac{1}{4}g_{\mu\nu}B^{\alpha\beta}B_{\alpha\beta}\right)\right.
\nonumber\\
\left.+g_{\mu\nu}\left(\frac{1}{4\pi}GV_\phi(\phi)+\frac{V_G(G)}{G^2}+\frac{V_\mu(\mu)}{\mu^2}+V_\omega(\omega)\right)\right]
\nonumber\\
+R_{\mu\nu}-\frac{1}{2}g_{\mu\nu}R+g_{\mu\nu}\Lambda
=-8\pi GT_{\mu\nu}.
\end{eqnarray}

\section{Static, Spherically Symmetric Vacuum Solution}
\label{sec:static}

In the static, spherically symmetric case with line element
\begin{equation}
ds^2=Bdt^2-Adr^2-r^2d\Omega^2,
\end{equation}
and with $d\Omega^2=d\theta^2+\sin^2{\theta}d\phi^2$, the field equations are written as
\begin{equation}
\frac{1}{A}\mu^2\phi_r+\frac{\partial V_\phi}{\partial\phi_r}=\frac{4\pi}{A\omega}J_r,
\end{equation}
\begin{eqnarray}
\phi_t''+\frac{2}{r}\phi_t'+\frac{\omega'}{\omega}\phi_t'+\frac{1}{2}\left(3\frac{A'}{A}-\frac{B'}{B}\right)\phi_t'
-A\mu^2\phi_t+AB\frac{\partial V_\phi}{\partial\phi_t}
\nonumber\\
=-\frac{4\pi A}{\omega}J_t,
\label{eq:phi}
\end{eqnarray}
\begin{eqnarray}
G''+\frac{2}{r}G'-\frac{3}{2}\frac{G'^2}{G}+\frac{1}{2}\left(\frac{\mu'^2}{\mu^2}-\omega'^2\right)G+\frac{1}{2}\left(\frac{B'}{B}-\frac{A'}{A}\right)G'&&\nonumber\\
+AV'_G(G)-3A\frac{V_G(G)}{G}-AG\left[\frac{V_\mu(\mu)}{\mu^2}+V_\omega(\omega)\right]-\frac{AG(R+2\Lambda)}{16\pi}
\nonumber\\
=0,
\end{eqnarray}
\begin{eqnarray}
\mu''+\frac{2}{r}\mu'-\frac{\mu'^2}{\mu}-\frac{G'}{G}\mu'+\frac{1}{4\pi}G\omega\left(\phi_r^2-\frac{A}{B}\phi_t^2\right)\mu^3
\nonumber\\
+\frac{1}{2}\left(\frac{B'}{B}-\frac{A'}{A}\right)\mu'-2A\frac{V_\mu(\mu)}{\mu}+AV'_\mu(\mu)=0,
\end{eqnarray}
\begin{eqnarray}
\omega''+\frac{2}{r}\omega'-\frac{G'}{G}\omega'+\frac{1}{8\pi}G\mu^2\left(\frac{A}{B}\phi_t^2-\phi_r^2\right)+\frac{1}{2}\left(\frac{B'}{B}-\frac{A'}{A}\right)\omega'
\nonumber\\
+\frac{1}{8\pi B}G\phi_t'^2-\frac{1}{4\pi}AGV_\phi(\phi)-AV'_\omega(\omega)=0,
\end{eqnarray}
\begin{eqnarray}
8\pi GT^t_t=-\Lambda-V-\frac{1}{A}N+\frac{A'}{A^2r}-\frac{1}{Ar^2}+\frac{1}{r^2}+\frac{G''}{AG}+\frac{2}{r}\frac{G'}{AG}
\nonumber\\
-2\frac{G'^2}{AG^2}-\frac{1}{2}\frac{A'G'}{A^2G}-\omega G\left(\frac{\phi_t'^2}{AB}+\frac{\mu^2\phi_t^2}{B}+\frac{\mu^2\phi_r^2}{A}\right),
\end{eqnarray}
\begin{eqnarray}
8\pi GT^r_r=-\Lambda-V+\frac{1}{A}N-\frac{B'}{ABr}-\frac{1}{Ar^2}+\frac{1}{r^2}+\frac{1}{2}\frac{B'G'}{ABG}+\frac{2}{r}\frac{G'}{AG}
\nonumber\\
-\omega G\left(\frac{\phi_t'^2}{AB}-\frac{\mu^2\phi_t^2}{B}-\frac{\mu^2\phi_r^2}{A}\right),
\end{eqnarray}
\begin{equation}
8\pi GT^t_r=-2\frac{G\omega\mu^2\phi_t\phi_r}{B},~~~~~~~~~8\pi GT^r_t=2\frac{G\omega\mu^2\phi_t\phi_r}{A},
\end{equation}
\begin{eqnarray}
8\pi GT^\theta_\theta=8\pi GT^\phi_\phi=-\Lambda-V-\frac{1}{A}N
+\frac{1}{2}\frac{A'}{A^2r}+\frac{1}{4}\frac{A'B'}{A^2B}
\nonumber\\
-\frac{1}{2}\frac{B'}{ABr}+\frac{1}{4}\frac{B'^2}{AB^2}
-\frac{1}{2}\frac{B''}{AB}+\frac{G''}{AG}+\frac{1}{r}\frac{G'}{AG}-2\frac{G'^2}{AG^2}-\frac{1}{2}\frac{A'G'}{A^2G}
\nonumber\\
+\frac{1}{2}\frac{B'G'}{ABG}
-\omega G\left(-\frac{\phi_t'^2}{AB}-\frac{\mu^2\phi_t^2}{B}+\frac{\mu^2\phi_r^2}{A}\right),
\end{eqnarray}
where
\begin{equation}
R=\frac{B''}{AB}-\frac{B'^2}{2AB^2}-\frac{A'B'}{2A^2B}+\frac{2B'}{ABr}-\frac{2A'}{A^2r}+\frac{2}{Ar^2}-\frac{2}{r^2},
\end{equation}
\begin{equation}
N=-4\pi\left(\frac{\mu'^2}{\mu^2}+\frac{G'^2}{G^2}-\omega'^2\right),
\end{equation}
\begin{equation}
V=2\omega GV_\phi(\phi)+8\pi\left[\frac{V_G(G)}{G^2}+\frac{V_\mu(\mu)}{\mu^2}+V_\omega(\omega)\right].
\end{equation}
The prime denotes differentiation with respect to $r$, i.e., $y'=dy/dr.$

These equations can be substantially simplified in the matter vacuum case ($T^\mu_\nu=0$), with no cosmological constant ($\Lambda=0$), setting the potentials to zero ($V_\phi=V_G=V_\mu, V_\omega=0$) and also setting $\phi_r=0$. These choices leave us with six equations in the six unknown functions $A$, $B$, $\phi_t$, $G$, $\mu$, and $\omega$:
\begin{eqnarray}
\frac{B'G'}{2ABG}-\frac{G'}{AGr}+2\omega G\left(\frac{\phi_t'^2}{AB}+\frac{\mu^2\phi_t^2}{B}\right)-\frac{B''}{2AB}+\frac{B'^2}{4AB^2}
\nonumber\\
+\frac{A'B'}{4A^2B}-\frac{B'}{2ABr}-\frac{A'}{2A^2r}+\frac{1}{Ar^2}-\frac{1}{r^2}=0,
\end{eqnarray}
\begin{eqnarray}
\frac{G''}{AG}-\frac{2G'^2}{AG^2}-\frac{B'G'}{2ABG}-\frac{A'G'}{2A^2G}+\frac{B'}{ABr}+\frac{A'}{A^2r}
\nonumber\\
+8\pi\left(\frac{G'^2}{AG^2}+\frac{\mu'^2}{A\mu^2}-\frac{\omega'^2}{A}-\frac{\omega G\mu^2\phi_t^2}{4\pi B}\right)=0,
\end{eqnarray}
\begin{eqnarray}
\omega G\left(\frac{\phi_t'^2}{AB}-\frac{\mu^2\phi_t^2}{B}\right)+4\pi\left(\frac{G'^2}{AG^2}+\frac{\mu'^2}{A\mu^2}-\frac{\omega'^2}{A}\right)
\nonumber\\
+\frac{B'G'}{2ABG}+\frac{2G'}{AGr}-\frac{B'}{ABr}-\frac{1}{Ar^2}+\frac{1}{r^2}=0,
\end{eqnarray}
\begin{equation}
\mu''+\frac{2}{r}\mu'-\frac{\mu'^2}{\mu}-\frac{G'}{G}\mu'+\frac{1}{2}\left(\frac{B'}{B}-\frac{A'}{A}\right)\mu'-\frac{A\omega G\phi_t^2}{4\pi B}\mu^3=0,\label{eq:mu}
\end{equation}
\begin{equation}
\omega''+\frac{2}{r}\omega'-\frac{G'}{G}\omega'+\frac{1}{2}\left(\frac{B'}{B}-\frac{A'}{A}\right)\omega'+\frac{G}{2B}\phi_t'^2+\frac{AG\mu^2\phi_t^2}{8\pi B}=0,\label{eq:omega}
\end{equation}
\begin{eqnarray}
G''+\frac{2}{r}G'-\frac{3}{2}\frac{G'^2}{G}+\frac{1}{2}\left(\frac{B'}{B}-\frac{A'}{A}\right)G'+\frac{1}{2}\left(\frac{\mu'^2}{\mu^2}-\omega'^2\right)G
\nonumber\\
-\frac{1}{16\pi}AGR=0,\label{eq:G}
\end{eqnarray}

The values of $A$, $B$, and $B'$ are fixed by the requirement that at large distance from a source, we must be able to mimic the Schwarzschild solution (albeit with a modified gravitational constant), and that at spatial infinity, the metric must be asymptotically Minkowskian. The vector field $\phi$ must also vanish at infinity, which provides another boundary condition. Next, we assume that the values of $G$, $\mu$, and $\omega$ are dependent on the source mass only, i.e., $G'=\mu'=\omega'=0$. We seek the remaining initial conditions in the form of the fifth force charge $Q_5$, and initial values of $G=G_0$, $\mu=\mu_0$, and $\omega=\omega_0$. We note that the basic properties of the numerical solution and the solution's stability are not affected by the values chosen for these parameters. However, their values must be chosen such that they correctly reflect specific physical situations. To determine these values, we now turn to the case of the point test particle.

\section{Test Particle Equation Of Motion}
\label{sec:testpart}

We begin by defining a test particle via its Lagrangian:
\begin{equation}
{\cal L}_\mathrm{TP}=-m+\alpha\omega q_5\phi_\mu u^\mu,
\label{eq:TP}
\end{equation}
where $m$ is the test particle mass, $\alpha$ is a factor representing the nonlinearity of the theory (to be determined later), $\omega$ is present as it determines the interaction strength, $q_5$ is the test particle's fifth-force charge, and $u^\mu=dx^\mu/ds$ is its four-velocity.

We assume that the test particle charge is proportional to its mass:
\begin{equation}
q_5=\kappa m,
\label{eq:kappa}
\end{equation}
with $\kappa$ constant and independent of $m$. This assumption implies that the fifth force charge $q_5$ is not conserved, as mass is not conserved. This is the case in Maxwell-Proca theory, as $\nabla^\mu J_\mu\ne 0$. We also have that the fifth force source charge $Q_5\propto M$.

From (\ref{eq:TP}), the equation of motion is obtained
\begin{equation}
m\left(\frac{du^\mu}{ds}+\Gamma^\mu_{\alpha\beta}u^\alpha u^\beta\right)=-\alpha\kappa\omega mB^\mu{}_\nu u^\nu.
\end{equation}
That $m$ cancels out of this equation is nothing less than a manifestation of the equivalence principle.

Our acceleration law can be written as~\cite{Toth}:
\begin{equation}
\ddot{r}=-\frac{G_NM}{r^2}\left[1+\alpha-\alpha(1+\mu r)e^{-\mu r}\right],
\label{eq:Yukawa}
\end{equation}
where $G_N$ is Newton's gravitational constant and $\alpha$ is given by
\begin{equation}
\alpha=\frac{M}{(\sqrt{M}+E)^2}\left(\frac{G_\infty}{G_N}-1\right),
\label{eq:alpha2}
\end{equation}
where $E$ is a constant of integration.

The acceleration law (\ref{eq:Yukawa}) can also be recast in the commonly used Yukawa form:
\begin{equation}
\ddot{r}=-\frac{G_YM}{r^2}\left[1+\alpha_Y\left(1+\frac{r}{\lambda}\right)e^{-r/\lambda}\right],
\end{equation}
with the Yukawa parameters $\alpha_Y$ and $\lambda$ given by
\begin{eqnarray}
G_Y&=&\frac{G_N}{1+\alpha_Y},\\
\alpha_Y&=&-\frac{(G_\infty-G_N)M}{(G_\infty-G_N)M+G_N(\sqrt{M}+E)^2},\\
\lambda&=&1/\mu=\frac{\sqrt{M}}{D}.
\end{eqnarray}
Here, $E$ and $D$ are two universal constants of integration which can be determined from fits to galaxy rotation curve data.

We can also express the acceleration law (\ref{eq:Yukawa}) as
\begin{equation}
\ddot{r}=-\frac{G_\mathrm{eff}M}{r^2},
\end{equation}
where the effective gravitational constant $G_\mathrm{eff}$ is defined as
\begin{equation}
G_\mathrm{eff}=G_N\left[1+\alpha-\alpha(1+\mu r)e^{-\mu r}\right].
\label{eq:Geff}
\end{equation}

\begin{figure}[t]
\includegraphics[width=0.8\linewidth]{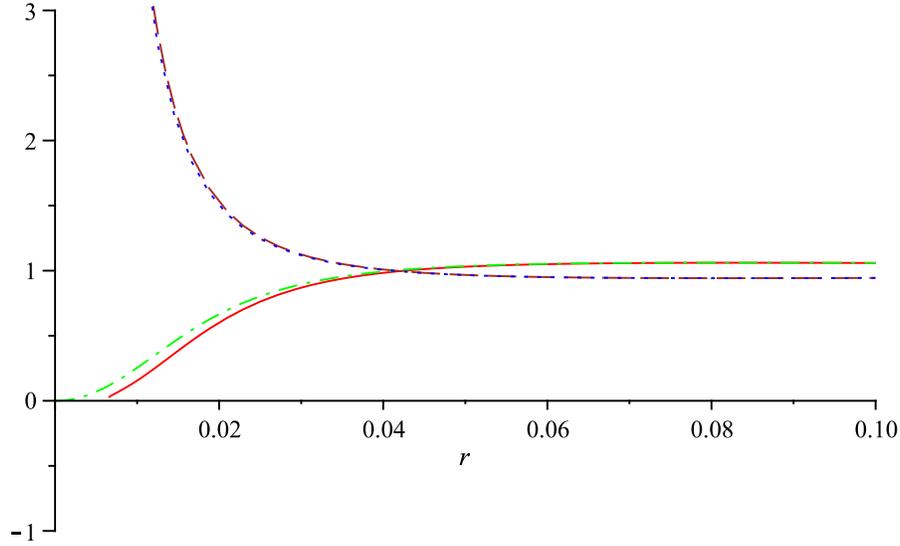}
\caption{Comparing MOG numerical solutions to the Reissner-Nordstr\"om solution, for a $10^{11}~M_\odot$ source mass. The MOG metric parameters $A$ (solid red line) and $B$ (dashed brown line) are plotted along with the Reissner-Nordstr\"om values of $A$ (dash-dot green line) and $B$ (dotted blue line). Horizontal axis is in pc. We observe that the $A$ metric parameter reaches 0 at below the Schwarzschild radius of a $10^{11}~M_\odot$ mass, which is $\sim 0.01$~pc.}
\label{fig:AB}
\end{figure}

The metric parameter $B(r)$ is given by
\begin{equation}
B(r)=1-\frac{2G_NM}{r}+\frac{(1+\alpha)G_N^2M^2}{r^2}.
\end{equation}
The $B(r)$ and $A(r)$ solutions are shown in Figure~\ref{fig:AB}.

\section{Cosmology}
\label{sec:cosmo}

In the case of a homogeneous, isotropic cosmology, using the Friedmann-Lema\^itre-Robertson-Walker (FLRW) line element,
\begin{equation}
ds^2=dt^2-a^2(t)[(1-kr^2)^{-1}dr^2+r^2d\Omega^2],
\end{equation}
the field equations assume the following form:
\begin{equation}
\ddot{\mu}+3H\dot{\mu}-\frac{\dot{\mu}^2}{\mu}-\frac{\dot{G}}{G}\dot{\mu}+\frac{1}{4\pi}G\omega\mu^3\phi_0^2+\frac{2}{\mu}V_\mu-V'_\mu=0,
\end{equation}
\begin{equation}
\ddot{\omega}+3H\dot{\omega}-\frac{\dot{G}}{G}\dot{\omega}-\frac{1}{8\pi}G\mu^2\phi_0^2+\frac{1}{4\pi}GV_\phi+V'_\omega=0,
\end{equation}
\begin{eqnarray}
\ddot{G}+3H\dot{G}-\frac{3}{2}\frac{\dot{G}^2}{G}+\frac{G}{2}\left(\frac{\dot{\mu}^2}{\mu^2}-\dot{\omega}^2\right)+\frac{3}{G}V_G-V_G'+G\left[\frac{V_\mu}{\mu^2}+V_\omega\right]
\nonumber\\
+\frac{G}{8\pi}\Lambda-\frac{3G}{8\pi}\left(\frac{\ddot{a}}{a}+H^2\right)=0,
\end{eqnarray}
\begin{eqnarray}
H^2+\frac{k}{a^2}=\frac{8\pi G\rho}{3}
-\frac{4\pi}{3}\left(\frac{\dot{G}^2}{G^2}+\frac{\dot{\mu}^2}{\mu^2}-\dot{\omega}^2-\frac{1}{4\pi}G\omega\mu^2\phi_0^2\right)
\nonumber\\
+\frac{2}{3}
\omega GV_\phi+\frac{8\pi}{3}\left(\frac{V_G}{G^2}+\frac{V_\mu}{\mu^2}+V_\omega
\right)
+\frac{\Lambda}{3}+H\frac{\dot{G}}{G},
\label{eq:FR1}
\end{eqnarray}
\begin{eqnarray}
\frac{\ddot{a}}{a}=-\frac{4\pi G}{3}(\rho+3p)
+\frac{8\pi}{3}\left(\frac{\dot{G}^2}{G^2}+\frac{\dot{\mu}^2}{\mu^2}-\dot{\omega}^2-\frac{1}{4\pi}G\omega\mu^2\phi_0^2\right)
\nonumber\\
+\frac{2}{3}
\omega GV_\phi+\frac{8\pi}{3}\left(\frac{V_G}{G^2}+\frac{V_\mu}{\mu^2}+V_\omega
\right)
+\frac{\Lambda}{3}+H\frac{\dot{G}}{2G}+\frac{\ddot{G}}{2G}-\frac{\dot{G}^2}{G^2},
\label{eq:FR2}
\end{eqnarray}
\begin{equation}
\omega\mu^2\phi_0-\omega\frac{\partial V_\phi}{\partial\phi_0}=4\pi J_0,~~~~~~~~~~J_i=0,
\end{equation}
where $H=\dot{a}/a$ is the Hubble expansion rate.

It is possible to obtain an exact numerical solution to this set of equations using numerical methods~\cite{Toth,Toth2}. To carry out the solution, we assume a pressureless matter equation of state $w=p/\rho=0$. Detailed fits to cosmological data, including the CMB angular power spectrum, the matter power spectrum and the SN1a supernovae data have been obtained~\cite{Toth2}. We find that the solutions can yield a ``bouncing'' cosmology. The bounce can be fine-tuned by choosing an appropriate value for $V_G$. This ensures that the universe reaches sufficient density in order to form a surface of last scattering. We emphasize that in our model only ordinary baryonic matter is present with a matter density of $\sim$5\% of the critical density. Nevertheless, the cosmology is flat, due in part to the increased value of the gravitational constant $G$, and in part to the presence of the non-zero energy density associated with $V_G$.

\section{Fitting Galaxy, Cluster Data And Solar System Data}
\label{sec:obs}

\begin{figure}[t]
\includegraphics[width=0.4\linewidth]{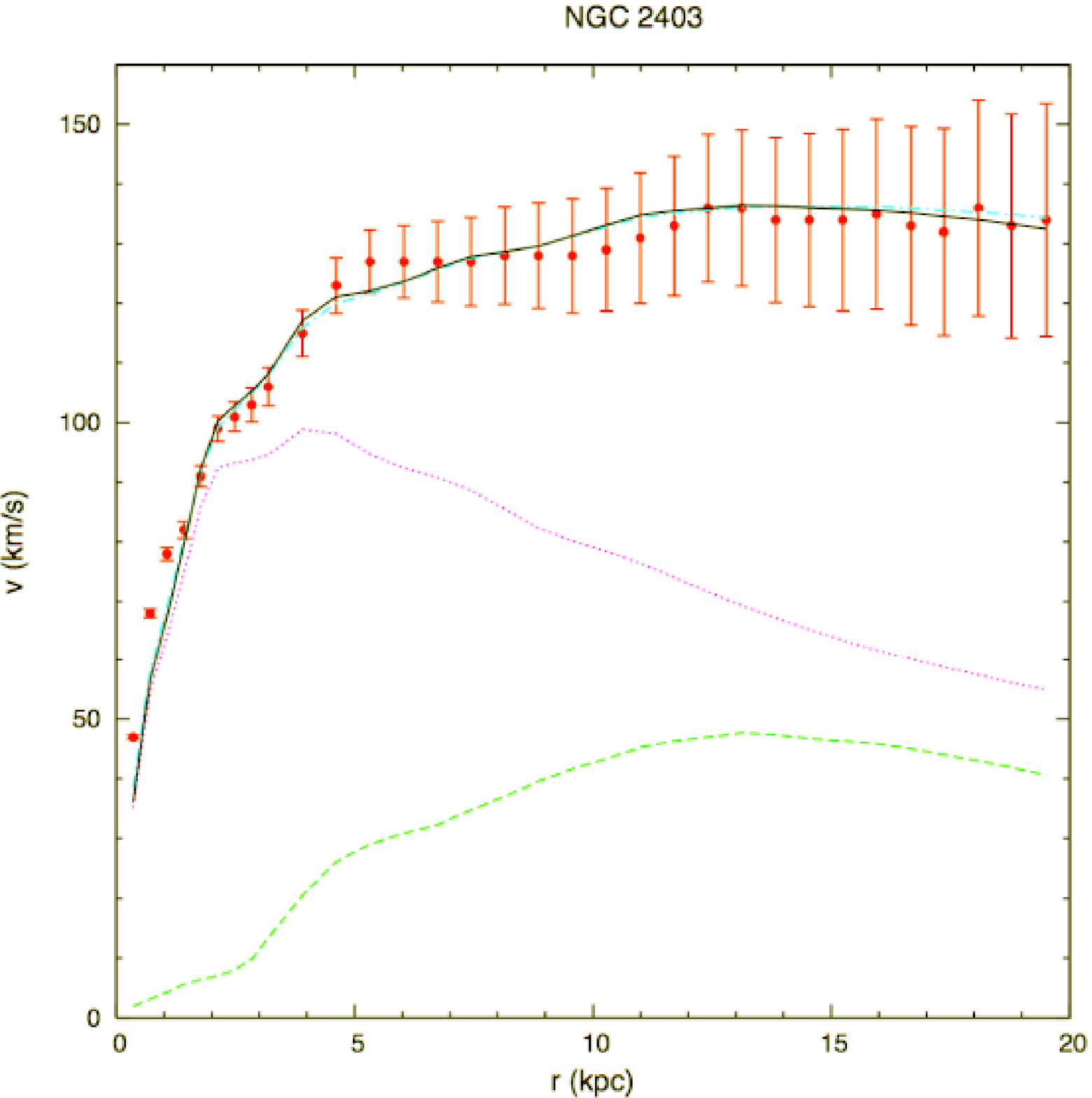}
\hskip 0.15\linewidth
\includegraphics[width=0.4\linewidth]{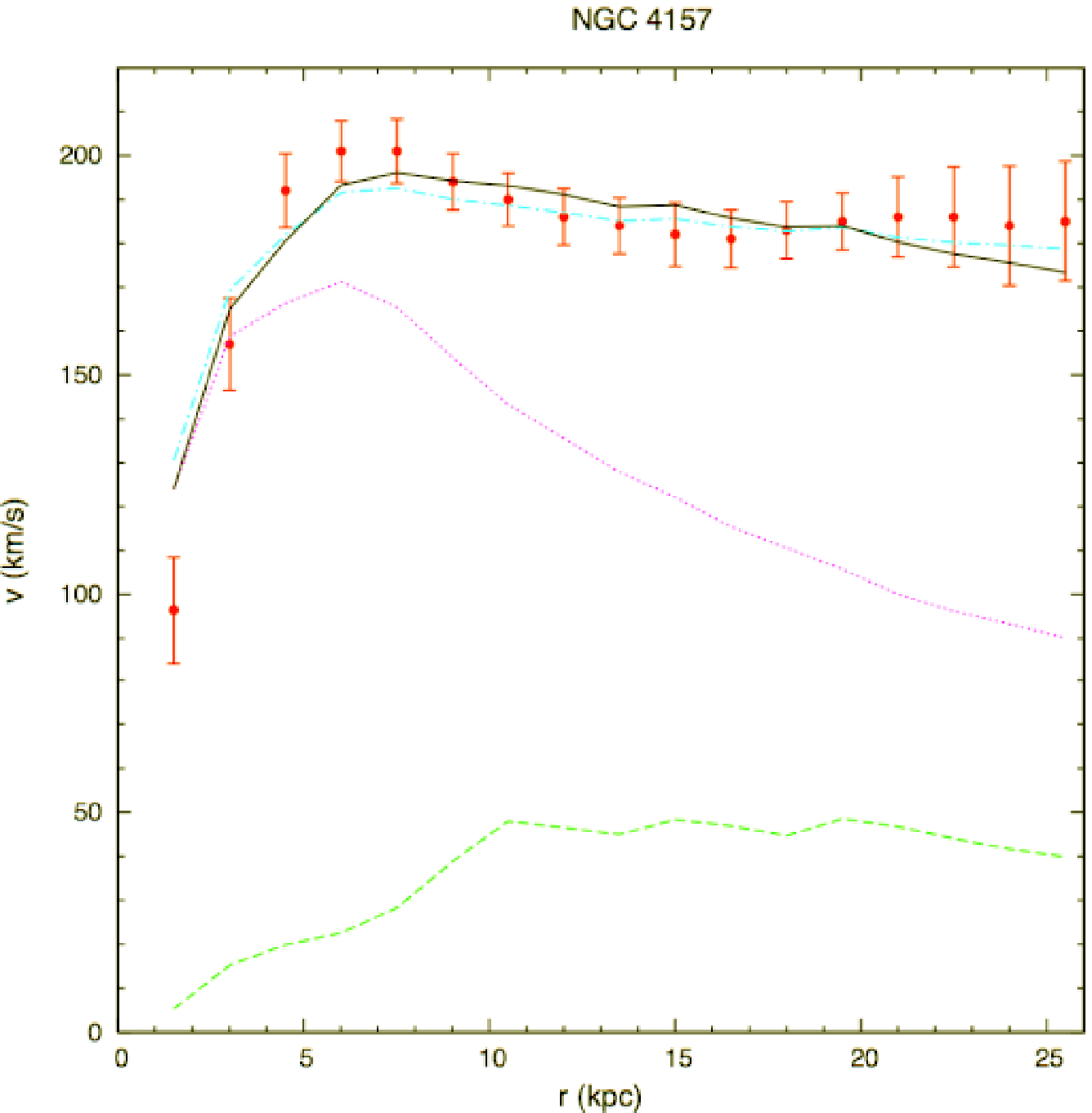}
\caption{Photometric fits to galaxy rotation curves. There are 2 benchmark galaxies presented here. Each is a best fit via the single parameter (M/L) based on the photometric data of the gaseous (HI plus He) and luminous stellar disks. The radial coordinate (horizontal axis) is given in kpc and the rotational velocity (vertical axis) in km/s. The red points with error bars are the observations, the solid black line is the rotation curve determined from MOG, and the dash-dotted cyan line is the rotation curve determined from MOND~\cite{Milgrom}. The other curves are the Newtonian rotation curves of the various separate components: the long-dashed green line is the rotation curve of the gaseous disk (HI plus He) and the dotted magenta curve is that of the luminous stellar disk (from \cite{Brownstein,Brownstein2}.}
\label{fig:gals}
\end{figure}

Unless one assumes that a massive dark matter halo is present, a typical spiral galaxy is dominated in mass by the central bulge. The motion of stars in the outer reaches of a galaxy can, therefore, be well approximated by the equations of motion in a static, spherically symmetric vacuum field. Indeed, our experience shows that the flat rotation curves of galaxies provide a sensitive test to determine the values of the constants $D$ and $E$. In particular, it is easy to see that our results so far are compatible with the Tully-Fisher law~\cite{Tully}.

Kepler's laws of orbital motion yield a relationship between circular orbital velocity $v_c$ at radius $r$ from a mass $M$ in the form
\begin{equation}
\frac{v_c^2}{r}=\frac{GM}{r^2}.
\label{eq:vc}
\end{equation}
Tully and Fisher~\cite{Tully} have determined that for galaxies, assuming that the brightness of a galaxy and its mass are correlated, the flat part of the rotation curve obeys the empirical relationship:
\begin{equation}
v_c^n\propto M,
\end{equation}
where $3\lesssim n\lesssim 4$. In our case, we obtain
\begin{equation}
v_c^2\propto\sqrt{M},
\end{equation}
corresponding to $n=4$ in the Tully-Fisher relationship.

Taking the next step, we select a small sample of galaxies and obtain an approximate fit to these galaxies yielding the values
\begin{eqnarray}
D&\simeq&6250~M_\odot^{1/2}\mathrm{kpc}^{-1},\label{eq:C2}\\
E&\simeq&25000~M_\odot^{1/2}.\label{eq:C1}
\end{eqnarray}

The galaxy rotation curves we obtain for galaxies of varying mass are in good agreement with these values, treating $D$ and $E$ as universal constants without dark matter (Figure~\ref{fig:gals}). The galaxy rotation curves were obtained modeling the galaxies as point masses, benefiting from photometric data, as in the more extensive fit to galaxy rotation velocities \cite{Brownstein,Brownstein2,Rodriguez}. This exercise demonstrates that our established relationships between $M$, $\alpha$, and $\mu$ not only satisfy the Tully-Fisher relationship, but also offer good agreement with actual observations. N-body simulations of galaxy rotation curve dynamics using MOG have been performed~\cite{Araujo}.

\begin{figure}[t]
\begin{tabular}{cc}
\includegraphics[width=0.38\linewidth]{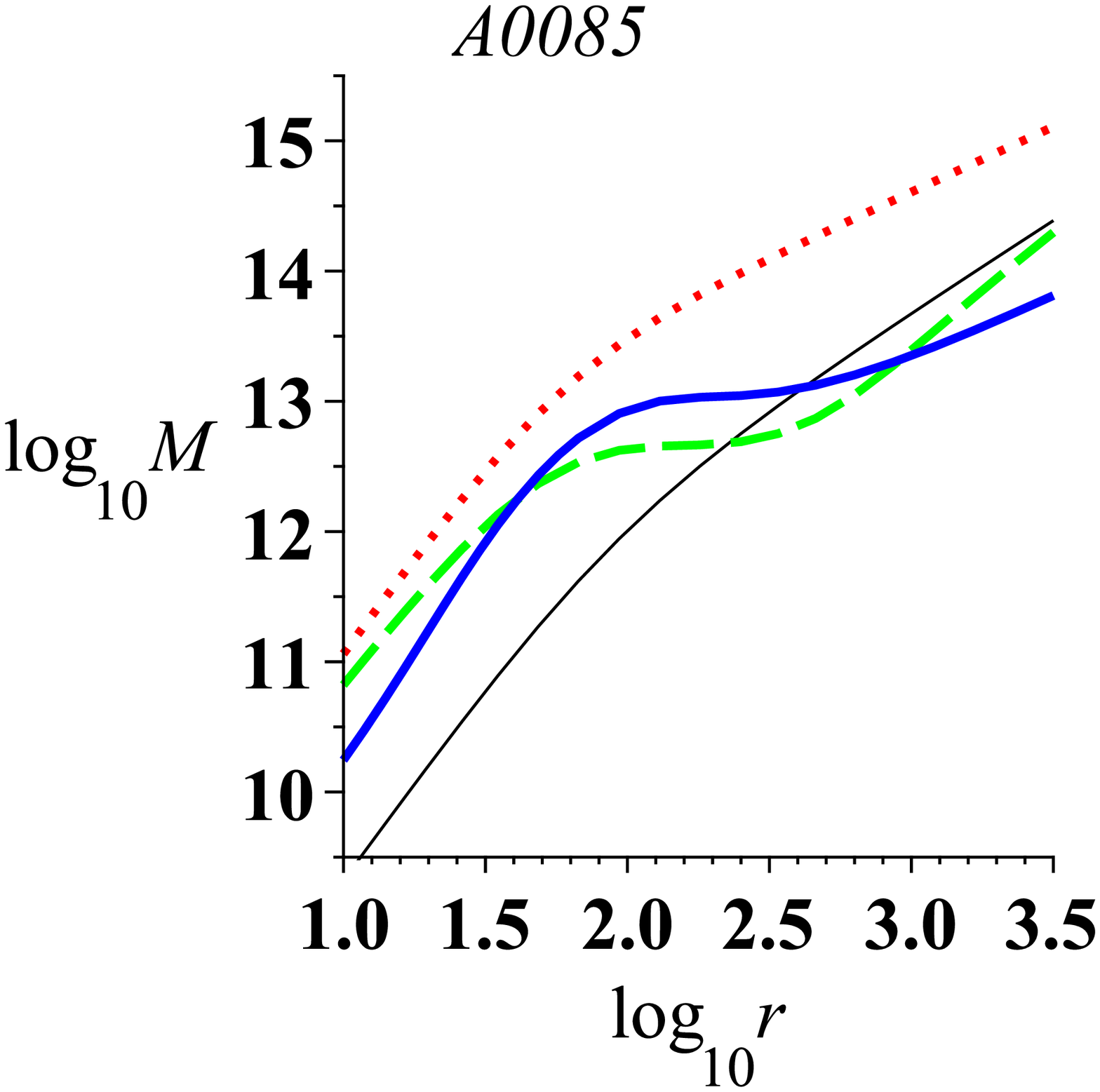}&\includegraphics[width=0.38\linewidth]{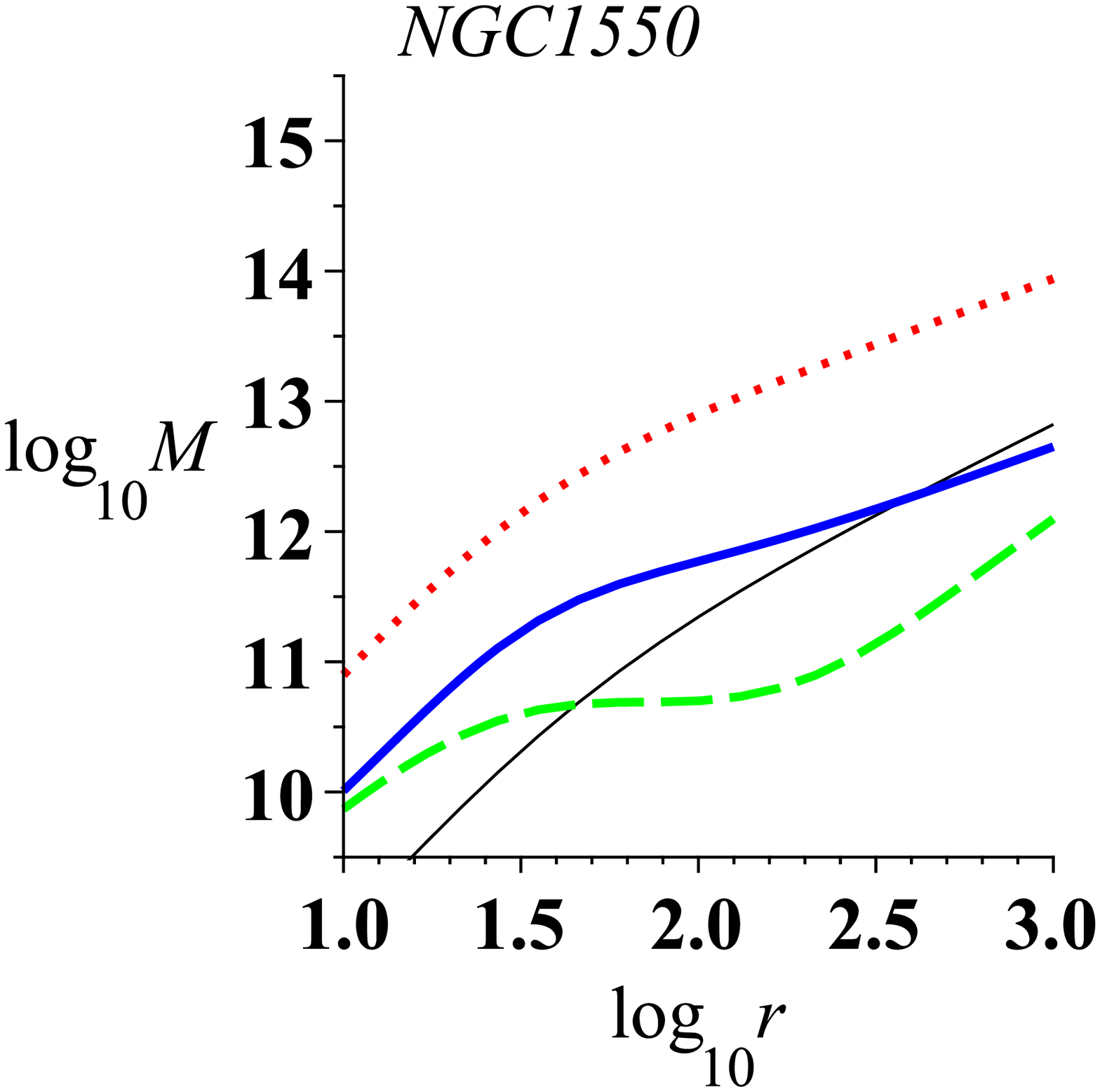}\\
\includegraphics[width=0.38\linewidth]{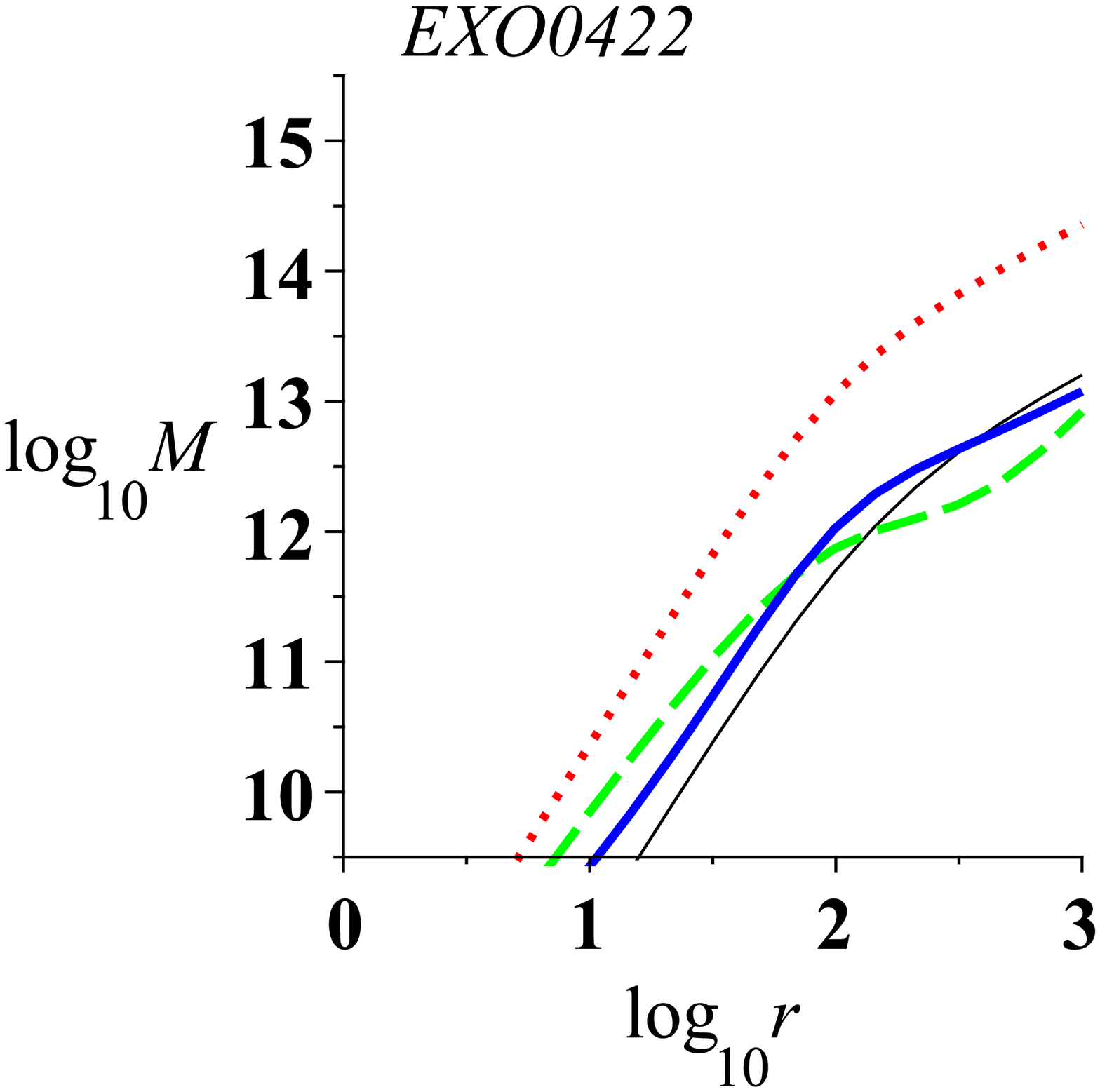}&\includegraphics[width=0.38\linewidth]{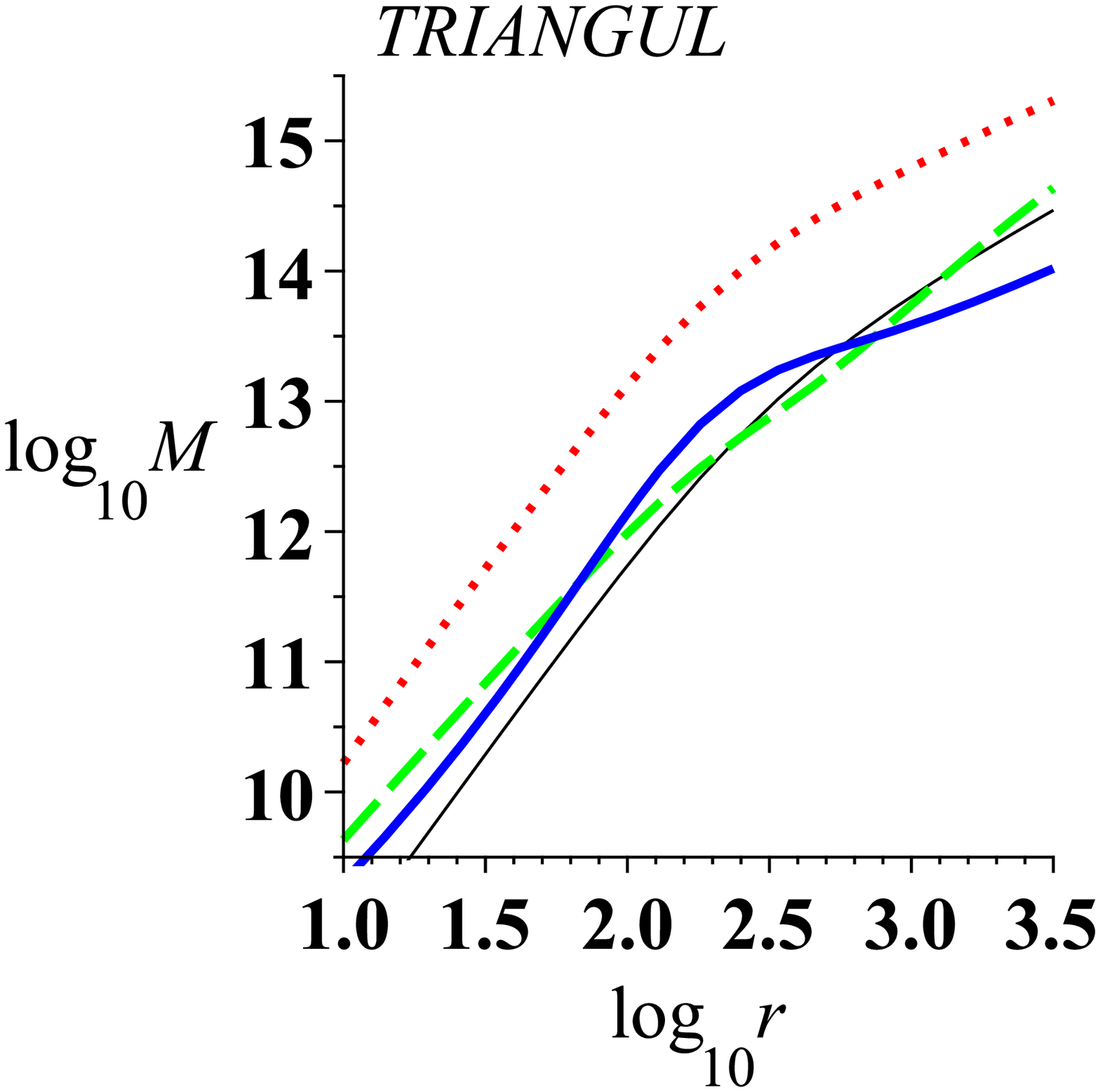}
\end{tabular}
\caption{A small sample of galaxy clusters studied in \cite{Brownstein2,Brownstein3}. Thin (black) solid line is the mass profile estimate from \cite{Reiprich}. Thick (blue) solid line is the mass profile estimated using our STVG results. Dashed (green) line is the result published in \cite{Brownstein3}, while the dotted (red) line is the Newtonian mass profile estimate. Radial distances are measured in kpc, masses in $M_\odot$.}
\label{fig:clusters}
\end{figure}

In \cite{Brownstein2,Brownstein3}, the spherically symmetric, static vacuum solution was used successfully to model galaxy clusters. We are able to produce a comparable result, while keeping the parameters $D$ and $E$ constant, by introducing an additional assumption: that the values of the MOG parameters $G_\infty$ and $\mu$ at some distance $r$ from the center of a spherically symmetric distribution of matter are determined not only by the amount of matter contained within radius $r$, but by the amount of matter within radius $r^*$. Figure~\ref{fig:clusters} shows the case of $r^*=3r$.

We have also succeeded in fitting the bullet cluster data~\cite{Clowe}, using MOG without dark matter~\cite{Brownstein4,Moffat2}.

We have applied MOG to predict dispersion velocity curves for globular clusters, and found that the predictions follow those of Newtonian gravity~\cite{Toth3}. By using Sloan Digital Sky Survey (SDSS) data, we have investigated how modified gravity theories including MOND and MOG affect satellite galaxies with the result that the data cannot currently differentiate significantly between modified gravity theories and dark matter models~\cite{Toth4}. The MOG prediction for lensing caused by intermediate galaxies and clusters of galaxies has been investigated~\cite{Toth5}.

\begin{figure}[t]
\includegraphics[width=0.48\linewidth,angle=270]{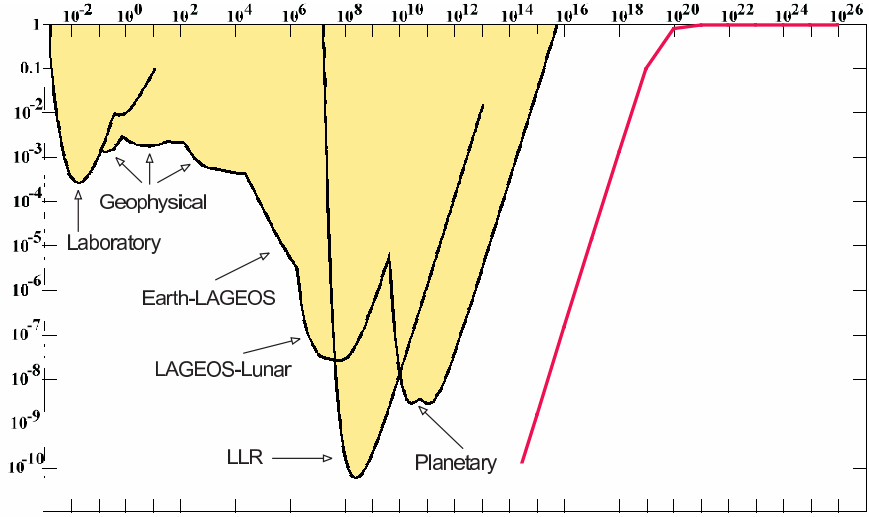}
\caption{Predictions of the Yukawa-parameters from the MOG field equations are not in violation of solar system and laboratory constraints. Predicted values of $\lambda$ (horizontal axis, in m) vs. $|\alpha_Y|$ are indicated by the solid red line. Plot adapted from \cite{Adelberger}.}
\label{fig:noses}
\end{figure}

The theory must also be consistent with experiments performed within the solar system or in Earthbound laboratories. Several studies (see, e.g., \cite{Adelberger}) have placed stringent limits on Yukawa-like modifications of gravity based on planetary observations, radar and laser ranging, and other gravity experiments. However, our prediction of the absolute value of the $\alpha_Y$ parameter is very small when $\lambda_Y$ is small. The latter is estimated at $\lambda_Y\simeq 0.16$~pc ($\sim 5\times 10^{15}$~m, or about 33,000~AU) for the Sun, and $\lambda_Y\simeq 2.8\times 10^{-4}$~pc ($\sim 8.7\times 10^{12}$~m, or $\sim 58$~AU) for the Earth. The corresponding values of $|\alpha_Y|$ are $|\alpha_Y|\simeq 3\times 10^{-8}$ and $|\alpha_Y|\simeq 9\times 10^{-14}$, respectively, clearly not in contradiction with even the most accurate experiments to date (Figure~\ref{fig:noses}).

In the solar system the MOG field equations become essentially those of the Jordan-Brans-Dicke model~\cite{Jordan,Brans}, for the influence of the vector field $\phi$ is reduced to very small values as shown in Figure~\ref{fig:noses}. However, the standard JBD model coupling constant $\omega_{\rm JBD}$ has to be  fine-tuned $\omega_{\rm JBD}> 40,000$ to fit the Cassini spacecraft measurement of the Eddington-Robertson, parameterized post-Newtonian parameter $\gamma-1=(2.1\pm 2.3)\times 10^{-5}$; the other parameter $\beta$ satisfies $\beta=1$ in MOG. We have resolved this problem in MOG by coupling the scalar field $G$ directly to matter by  means of a scalar matter current:
\begin{equation}
J=-\frac{1}{2}GT,
\end{equation}
where $T={T^\mu}_\mu$. This leads to obtaining an agreement with Earth based equivalence principle experiments and $\gamma=1$~\cite{MoffToth}.

\begin{figure}[t]
\includegraphics[width=0.8\linewidth]{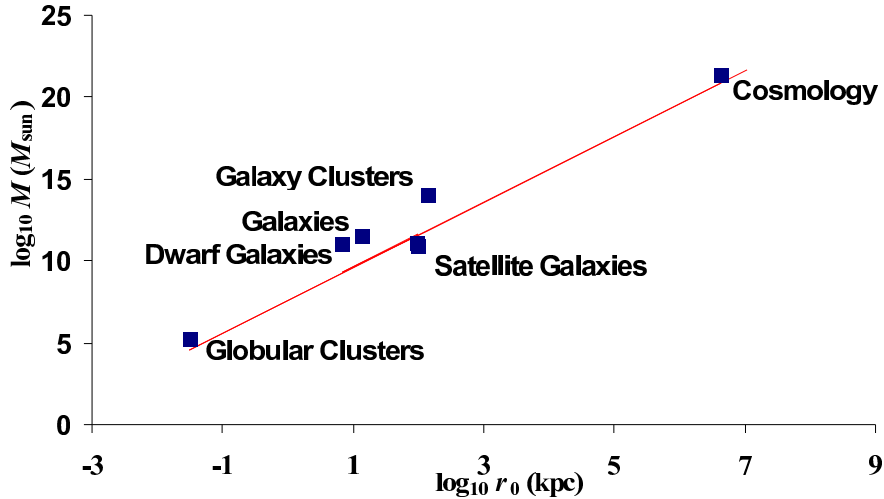}
\caption{The relationship $\mu^2M=$~const. between mass $M$ and the Yukawa-parameter $r_0=\mu^{-1}$ across many orders of magnitude remains valid. The solid red line represents our theoretical prediction.}
\label{fig:Mr}
\end{figure}

We have plotted $M$ vs. $r_0=\mu^{-1}$ in Figure~\ref{fig:Mr}. For the purposes of this plot, we used previously published results, while noting that our new calculations place dwarf galaxies, galaxies, and galaxy clusters by definition exactly on the line representing our prediction. This plot demonstrates the validity of MOG from the scales of star clusters to cosmological scales.

We have investigated the possibility that MOG can explain in a fundamental way the origin of inertial mass. The static, spherically symmetric solution does not satisfy Birkhoff's theorem as in the case of the Schwarzschild solution in GR. This leads to a Mach-type influence of distant matter that can determine the inertial mass of a body. A possible spacecraft experiment has been proposed to test this prediction~\cite{Toth6}.

On the scale of Earth-based laboratory and solar system experiments with ever greater precision, MOG may eventually be verified or falsified. Beyond the solar system, as larger galaxy samples become available, the presence or absence of baryonic oscillations in the matter power spectrum may unambiguously decide in favor of modified gravity theories or dark matter~\cite{Toth,Toth2}. Confirmed detection of dark matter particles in deep space or in the laboratory would also be a strong indication against modified gravity.

\section{Conclusions}
\label{sec:concl}

In this paper, we have demonstrated how results of our Modified Gravity (MOG) theory can be derived directly from the action principle, without resorting to the use of fitted parameters. After we fix the values of some integration constants from observations, no free adjustable parameters remain, yet the theory remains consistent with observational data in the two cases that we examined: the vacuum solution of a static, spherically symmetric gravitational field, and a cosmological solution. These solutions were explored using numerical methods, avoiding the necessity to drop terms or make other simplifying assumptions in order to obtain an analytic solution. Further, the constraints used to compute the solutions are consistent to the extent that they overlap with one another. The fact that at the level of the calculations presented here, our theory is not obviously falsified is an indication that we should pursue MOG further, for instance by obtaining interior solutions to the MOG field equations, and using these solutions to develop tools to perform $N$-body simulations.

\section*{Acknowledgments}

The research was partially supported by National Research Council of Canada. Research at the Perimeter Institute for Theoretical Physics is supported by the Government of Canada through NSERC and by the Province of Ontario through the Ministry of Research and Innovation (MRI).


\end{document}